# Hysteretic Phonons and Quasielastic Response: A Raman Study of Thermal Memory in Two-dimensional CuCrP$_2$S$_6$

Chaitanya B. Auti[1, *], Atul G. Chakkar[1], Sebastian Selter[2], Yuliia Shemerliuk[2], Bernd Büchner[2,3], Saicharan Aswartham[2], and Pradeep Kumar[1#]

[1]*School of Physical Sciences, Indian Institute of Technology, Mandi-175005, India.*
[2]*Institute for Solid State Research, Leibniz IFW Dresden, Helmholtzstr. 20, 01069 Dresden, Germany*
[3]*Institute of Solid State and Materials Physics and Würzburg-Dresden Cluster of Excellence ct.qmat, Technische Universität Dresden, 01062 Dresden, Germany*

**Abstract**

We present a comprehensive temperature and polarization dependent inelastic light scattering (Raman) study on single crystals of two-dimensional CuCrP$_2$S$_6$, a layered van der Waals material exhibiting coupled magnetic and electric degrees of freedom. Raman measurements were performed from 5 to 300 K to probe phonon dynamics across multiple structural and magnetic phase transitions. Our analysis reveals pronounced thermal hysteresis in phonon self-energy parameters and dynamic Raman susceptibility, confirming the first-order nature of the antipolar transition near $T_{C1}$ ~ 145 K and a second-order transition near $T_{C2}$ ~ 190 K. Low-frequency modes associated with Cu$^+$ and Cr$^{3+}$ ions exhibit softening and anomalous linewidth behaviour, in particular phonon mode P2 (~ 37 cm$^{-1}$) showing non-monotonic temperature dependence and intensity enhancement near 60 K suggesting persistent off-centre Cu$^+$ dynamics in the quasi-antipolar phase. The coexistence and coupling of soft phonon modes and central peaks indicate a crossover from displacive to order-disorder type transition mechanisms. Additionally, phonon anomalies below the Néel temperature ($T_N$ ~ 32 K) reflect spin-phonon coupling, linking lattice vibrations to long-range magnetic correlations. Our findings provide critical insight into the lattice instabilities, symmetry evolution, and quasiparticle interactions in CuCrP$_2$S$_6$, offering a deeper understanding of phase transition dynamics in two-dimensional multiferroic systems and guiding future design of magnetoelectric and spintronic devices.

*autichaitanya5@gmail.com
#pkumar@iitmandi.ac.in

## 1. Introduction

In recent years, materials exhibiting multiferroicity - characterized by the coexistence of ferro/antiferromagnetic and ferro/antiferroelectric ordering have attracted considerable scientific interest. This surge is driven by their potential in cutting-edge applications such as spintronics, magnetoelectric memory devices, and multifunctional electronic systems [1,2]. However, conventional oxide-based multiferroics face persistent challenges, including pronounced size effects, dangling bonds, and interfacial instabilities. These limitations impede device miniaturization and restrict their broader technological integration [3]. Two-dimensional (2D) van der Waals (vdW) materials exhibit unique physical properties that distinguish them from conventional systems. Notably, they possess dangling bond-free surfaces [4,5] and stable layered architectures characterized by strong intralayer bonding and weak interlayer interactions, features that facilitate mechanical exfoliation. These 2D sheets and their heterostructures hold immense promise for next-generation electronic devices [6]. Among them, the heterocharge metal thiophosphate (MTP) family, with the general formula $M^{I}M^{III}[P_2S_6]^{4-}$, (where $M^{I}$ = $Ag^+$, $Cu^+$, $M^{III}$ = $In^{3+}$, $V^{3+}$, $Sc^{3+}$, $Cr^{3+}$), has garnered particular attention. This is due to its vdW-layered crystal structure combined with tunable magnetic and electric properties governed by the choice of transition metal cations.

Among these MTPs, $CuInP_2S_6$ (CIPS) stands out as a widely investigated ferroelectric material, known for exhibiting room-temperature ferroelectricity. This behaviour arises from the opposite displacement of $Cu^+$ and $In^{3+}$ ions, forming a dual sublattice that induces net polarization below the critical temperature ($T_C$) of ~ 315 K. Due to the absence of unoccupied $d$ orbitals in both $Cu^+$ and $In^{3+}$, no magnetic ordering is expected in CIPS [7]. Intriguingly, a closely related compound CuCrP$_2$S$_6$ (CCPS) demonstrates markedly different behaviour upon substituting In$^{3+}$ with Cr$^{3+}$. CCPS exhibits magnetic ordering below the Néel temperature $T_N \sim$ 32 K, along with antipolar ordering below $T_{C1} \sim$ 145 K and a quasi-antipolar phase below $T_{C2} \sim$ 190 K [8]. Recent investigations on few-layer CCPS reveal a ferroelectric-paraelectric transition near 333 K [6], while second harmonic generation studies suggest a higher Curie temperature of approximately 470 K, beyond which ferroelectric domains are completely suppressed [9].

The investigation of first-order phase transitions in anti- and ferroelectric materials reveals a complex interplay of microscopic mechanisms that shape their structural and dielectric behaviour near the transition temperature. A hallmark of such transitions is thermal hysteresis, evident in both phonon spectra and dielectric measurements, which signals a discontinuous change in the order parameter [10]. The presence of thermal hysteresis, often associated with first-order transitions, underpins non-volatile memory behaviour, enabling bistable switching critical for applications in ferroelectric RAM and energy-efficient logic devices [11-13]. Two distinct dynamical signatures are commonly employed to characterize the nature of these transitions: central peaks indicative of order-disorder behaviour, as observed in NaNO$_2$, NH$_4$Cl [14,15], and low-frequency soft phonon modes typical of displacive transitions, seen in materials such as SrTiO$_3$ [16], BaTiO$_3$ [17], and SbSI [18]. Remarkably, in many complex anti/ferroelectrics, both features coexist near the transition point, suggesting a crossover between displacive and order-disorder mechanisms. This duality has been reported in materials such as Cd$_2$Nb$_2$O$_7$ [19], Gd$_2$(MoO$_4$)$_3$ [20], Sn$_2$P$_2$Se$_6$ [21], and PbHfO$_3$ [22]. Deciphering this crossover is essential for advancing our understanding of first-order transitions. It not only deepens insight into lattice dynamics but also offers a pathway to engineer materials with tailored phase transition characteristics, thermal responses, and multifunctionality [20,21].

While CCPS has been the subject of investigations including neutron diffraction [23] and temperature and pressure dependent Raman spectroscopy studies [24,25]; however, its low-frequency Raman scattering characteristics remain largely unexplored, particularly in relation to thermal hysteresis. Although considerable attention has been given to the temperature dependence of higher-symmetry phonon modes in the *C2/c* phase at room temperature, the behaviour of lower-symmetry phonon modes at low temperature in the *Pc* space group remains largely unexplored. Raman spectroscopy serves as a powerful, non-destructive tool for probing quasiparticle excitations in two-dimensional materials. It provides direct access to phonon dynamics, electron-phonon coupling, and collective excitations that underpin the functional properties of these systems [26-28]. In anti/ferroelectric materials, Raman scattering is especially effective in revealing soft-mode behaviour, phonon anomalies, and central peaks-key signatures of both first- and second-order phase transitions [29,30].

In this work, we present a comprehensive temperature and polarization dependent Raman scattering study of single crystals of CCPS, encompassing both higher and lower-symmetry phonon modes. Our measurements reveal pronounced thermal hysteresis in phonon frequencies and dynamic Raman susceptibility, indicative of first-order phase transition behaviour. Multiple phase transitions are identified through distinct anomalies in the self-energy parameters of the phonon modes. Additionally, we probe the low-frequency quasi-elastic scattering and soft-mode dynamics, which further illuminate the underlying lattice instabilities. The polarization-dependent intensity profiles of phonon modes provide insight into their symmetry characteristics and crystal structure. This study aims to unravel key features of first-order phase transitions such as thermal hysteresis, structural transformations, and phonon anomalies; while addressing the long-standing ambiguity between displacive and order-disorder mechanisms governing the transition.

**2. Experimental details**

Single-crystal samples of CCPS were synthesized via the chemical vapor transport method, as described in the reference [31]. Temperature-dependent Raman scattering measurements were carried out using a LabRAM HR-Evolution spectrometer in the backscattering geometry, integrated with a closed-cycle refrigerator (Montana) to achieve a temperature range of 5-300 K with a precision of ±0.1 K. Spectra were excited using a 633 nm laser, with the incident power maintained below 0.5 mW to minimize the local heating effects. A long working distance objective (50X, NA = 0.5) was employed for both focusing on the sample surface and collecting the scattered light, which was detected using a Peltier-cooled charge-coupled device. Raman measurements were done upon initial heating of the sample from 5 K to 300 K (termed as heating cycle) and subsequent cooling of the sample from 300 K to back to 5 K (termed as cooling cycle). To resolve the symmetry characteristics of the phonon modes, polarization-resolved Raman measurements were performed using a half-wave plate and analyser. These measurements were conducted at two temperatures i.e. 5 K and 300 K under both parallel and perpendicular polarization configurations.

## 3. Results and discussion

### 3.1 Crystal structure and phonon modes

At room temperature, CCPS crystallizes in the monoclinic *C2/c* space group (point group $C_{2h}$), identical to that of CIPS above its ferroelectric transition temperature ($T_C$ ~ 315 K). The primitive unit cell of CCPS comprises 20 atoms, yielding 60 phonon modes, including 3 acoustic modes and 33 Raman-active modes i.e. $16Ag + 17Bg$. Among these, the low-frequency mode near ~ 56 cm$^{-1}$, labelled as T(Cr), corresponds to the translational motion of Cr$^{3+}$ ions. Four internal modes labelled as R(PS$_3$), T(PS$_3$), $v$(P-P), and $v$(P-S) in Fig. 1c-(i) (denoted as P25, P35, P46, and P56 in the text), are associated with rotational, translational, and stretching deformations of the P$_2$S$_4^-$ sublattice, respectively [32]. Upon cooling below $T_{C1}$ ~ 145 K, the crystal symmetry lowers to the *Pc* space group (point group $Cs$), with a doubled unit cell containing 40 atoms. This results in 120 phonon modes, including 3 acoustic modes and 117 Raman-active modes (59 $A'$ + 58 $A''$). The low-temperature phase of CCPS features quasi-trigonal $CuS_3$ units, octahedral $CrS_6$ coordination, and triangular networks formed by $P_2S_6$ clusters [24]. Cr ions and P-P pairs occupy nearly central positions within the layers, and the $Cr^{3+}$ ions (spin $S = 3/2$) introduce magnetic ordering. $Cu^+$ cations adopt a nearly trigonal planar coordination, with their positions significantly displaced toward either the top or bottom sulphur layers (see Supplementary Material, Fig. S1). This displacement breaks inversion symmetry and induces a local electric dipole moment. Below the Néel temperature, $Cr^{3+}$ ions within each layer are ferromagnetically coupled, while interlayer coupling is antiferromagnetic (AFM) [31]. At room temperature, the $Cu^+$ ions randomly occupy upper and lower sites, leading to a net paraelectric state.

In CCPS, vibrational modes are classified based on their frequency range and the atomic species involved. Recent phonon projected density of states calculations reveals distinct contributions from individual atoms [33]. Low-frequency modes (below ~ 150 $cm^{-1}$) predominantly arise from extended lattice vibrations involving heavier metal cations, particularly $Cu^+$ ions. Notably, the mode P2 at ~ 37 $cm^{-1}$, labelled as T(Cu), corresponds to the translational motion of $Cu^+$, and P3 at ~ 72 $cm^{-1}$ are identified as a soft mode's indicative of lattice instability. Chromium ($Cr^{3+}$) ions contribute primarily to mid-frequency vibrations spanning ~ 50-360 $cm^{-1}$, reflecting their intermediate mass and bonding environment. In contrast, phosphorus (P) atoms are responsible for high-frequency phonon modes in the ~ 500 - 620 $cm^{-1}$ range. Sulphur (S) atoms play a pivotal role across the spectrum, exhibiting a broad frequency distribution due to their direct coordination with Cu, Cr, and P atoms. This diverse bonding landscape makes S a key mediator in phonon dispersion and lattice dynamics.

Figure 1 presents the unpolarized Raman spectrum of CCPS recorded at 5 K using 633 nm laser excitation, covering a spectral range from ~ 14 to 700 cm$^{-1}$. The self-energy parameters of the phonon modes i.e. peak frequency (ω), full width at half maximum (FWHM), and the intensity were extracted by fitting the spectra with a sum of Lorentzian functions (see Supplementary Material, Fig. S2 for fitted spectrum). At room temperature, eight phonon modes (P25, P35, P39, P46, P52, P53, P56, and P58) were observed, whereas at low temperature, 58 distinct modes (P1-P58) were identified (see Supplementary Material, Fig. S5 for temperature evolution). A sudden appearance of multiple peaks just below $T_{C1}$ (~ 145 K) strongly supports the first-order nature of the antipolar phase transition. The tabulated phonon frequencies at 5 K and 300 K are given in Table 1 of the Supplementary Material. The Raman spectra of CCPS reveal several notable phenomena: (i) thermal hysteresis in the self-energy parameters of phonon modes, (ii) the merging of soft phonon modes into a quasielastic scattering peak, and (iii) thermal hysteresis in the dynamic Raman susceptibility. These observations are discussed in detail in the following sections.

### 3.2 Temperature dependence of the phonon modes

Figures 1(a-b) present 2D colour contour plots of the Raman spectra of CCPS across the temperature ranges 5-300 K. Distinct intensity variations in these plots clearly mark the underlying phase transitions. Figures 1(c-i-iii) display representative spectra from the heating and cooling cycles: the spectrum at 5 K corresponds to the combined AFM and antipolar phase; the 90 K spectrum captures the onset of the transition from antipolar to quasi-antipolar phase; and the 290 K spectrum reflects the paraelectric phase. Figure 2 illustrates the temperature dependence of the phonon frequency (ω), linewidth (FWHM), and normalized intensity for selected modes P2, P12, P34, and P36. Additional phonon mode data are provided in Supplementary Material (Figs. S3 and S4). These modes, associated with the low-symmetry $P_c$

phase, disappear near the antipolar transition temperature ($T_{C1} \sim 145$ K), highlighting their critical role in the structural transition of CCPS. Their abrupt changes further confirm the first-order nature of the phase transition.

Soft phonon modes are dynamically unstable lattice vibrations whose frequencies approach zero near a phase transition, signalling an impending structural reconfiguration. In CCPS, we observed softening of modes P2 and P3 by approximately 2 cm$^{-1}$ and 10 cm$^{-1}$, respectively; upon heating (see Supplementary Material, Fig. S3). This behaviour is indicative of soft-mode dynamics and can be interpreted within the framework of mean-field theory, $\omega(T) = \beta(T^* - T)^{1/2} + \omega_0$, where $T^*$ is the characteristic temperature at which a mode vanishes and $\beta$ is constant. Here the phonon frequency follows a temperature-dependent relation governed by a characteristic transition temperature and a non-zero offset term $\omega_0$ [24]. Unlike the complete mode condensation expected in displacive-type transitions, the soft modes in CCPS retain finite frequencies across the transition, with $\omega_0$ remaining non-zero. The non-zero constant term $\omega_0$ signifies a critical distinction that supports the order-disorder type structural transition responsible for the development of defect dipoles associated with the Cu ions. Several factors may contribute to the saturation of soft-mode softening such as (i) first-order nature of the structural phase transition, which inherently limits continuous mode evolution. (ii) Quantum effects at low temperatures, as observed in incipient ferroelectrics like SrTiO$_3$ and KTaO$_3$. (iii) Non-criticality of the polar soft mode, where the actual soft mode driving the transition originates away from the Brillouin zone centre - typical of antiferroelectric, incommensurate, or improper ferroelectric transitions. (iv) Coupling with central modes, which exhibit enhanced dispersion near $T_{C1}$ and lie below the soft-mode frequency. The central mode dominates the dielectric response and are responsible for the anomaly near the transition temperature [19,30]. In CCPS, the observed merging of soft phonon modes into a central mode near $T_{C1}$ suggest a crossover from displacive to order-disorder type transition dynamics.

Phonon modes P2, P34, and P36 exhibit clear softening upon heating, a trend similarly observed in modes P3, P16, and P44 (see Supplementary Material, Fig. S3). FWHM of most modes increases with increasing temperature, consistent with enhanced phonon-phonon scattering. However, mode P2 displays an anomalous where its FWHM decreases up to 15 K, follows the expected trend until ~ 80 K, and then begins to decrease again. The intensity of mode P2 increases with increasing temperature during the heating cycle, peaking around 60 K before declining, suggesting that lattice dynamics associated with off-centre $Cu^+$ cations remain active even in the quasi-antipolar phase. Modes P12 and P36 show a monotonic decrease in intensity with increasing temperature, a trend also observed for modes P16, P33, and P44 (see Supplementary Material Fig. S3). In contrast, mode P34 exhibits a sudden intensity enhancement beyond 70 K, while P3 shows a steady increase with increasing temperature. Notably, phonon modes P2, P34, and P36 (see Fig. 2), along with P25, P35, P46, and P56 (see Fig. S4), display a distinct change in frequency slope and gradual softening below ~ 30 K, coinciding with the AFM ordering temperature. Both FWHM and intensity profiles of these modes exhibit slope changes near $T_N$, pointing to anomalous phonon behaviour attributed to the spin-phonon coupling. This coupling remains significant as long as long-range magnetic correlations persist, underscoring the interplay between lattice vibrations and magnetic ordering in CCPS [34,35].

Nearly all higher-symmetry phonon modes in CCPS exhibit clear signatures of the underlying phase transitions, as reflected in their self-energy parameters. Notably, modes P25 and P35, associated with substantial displacements of metal atoms i.e. $Cu^+$ and $Cr^{3+}$ ions, display pronounced deviations from quasi-harmonic behavior due to the large amplitude of atomic motion [33]. These deviations underscore the anharmonic nature of lattice dynamics near the transition points. Phonon modes presented in Fig. S4 consistently reveal features characteristic of the first-order antipolar phase transition around $T_{C1} \sim 140$ K. Additionally, modes P25, P35, P39, and P56 show distinct anomalies near $T_{C2} \sim 190$ K, indicative of a second-order quasi-antipolar phase transition. These observations highlight the sensitivity of specific phonon modes to both structural and symmetry changes, offering valuable insight into the complex phase evolution in CCPS.

### 3.3 Thermal hysteresis in phonon modes

Upon initial heating from 5 K to 300 K and subsequent cooling of the sample from 300 K to back to 5 K, the phonon mode intensities fail to fully recover, indicating an incomplete reversibility across the thermal cycle (see Supplementary Material, Fig. S1). This behaviour reflects a pronounced thermal hysteresis in the onset of the structural phase transition. As shown in Figure 3, the Raman spectra recorded at 90 K, 80 K, and 70 K reveal that CCPS has transitioned back to the antipolar phase upon cooling; however, several characteristic peaks remain absent. Generally, a first-order phase transition necessitates a certain degree of overheating (undercooling) above (below) the equilibrium transition temperature, where the free-energy curves of the two phases intersect with a discontinuity in the first derivative. The system stabilizes in one of the two (meta)stable states depending on its thermal trajectory, i.e., whether the temperature is ramping up or down [36,37].

A pronounced discontinuity in phonon behaviour is observed near the first-order phase transition temperature ($Tc_1 \sim 145$ K). Upon heating, several phonon modes abruptly vanish, and their reappearance during cooling occurs at distinctly lower temperatures, revealing a clear hysteresis in the self-energy parameters, a hallmark of first-order transitions. As illustrated in Figure 2, the initial heating of the sample leads to a decrease in phonon mode frequencies. However, the cooling trajectory does not retrace the frequency-temperature curve obtained during heating, indicating irreversible dynamics. Soft modes P2 and P3 exhibit substantial hysteresis, with their frequencies decrease upon heating and subsequently increase during cooling, forming a closed hysteresis loop (see Supplementary Material Fig. S3). Mode P12 similarly displays a complete hysteresis loop, while modes P34 and P36 show asymmetric hysteresis behaviour. Notably, the intensity of mode P2 increases with increasing temperature during heating, peaking around 60 K, and then diminishes. In contrast, during cooling, its intensity rises until ~ 30 K before declining again. It is evident that a ~ 20-30 K thermal delay causes phonon modes to reappear as the system cools. Figure 4 highlights phonon modes that persist to higher temperatures and belong to the higher-symmetry point group (*C2/c*). These modes i.e. P35, P52, and P58 exhibit significant hysteresis in their self-energy parameters in the temperature range of 30-160 K. The frequency-temperature curve for heating from 60 K to 150 K remains consistently lower than that of the cooling cycle, with the exception of mode P58. Above 150 K, both curves converge, marking the paraelectric-antiferroelectric boundary. Additionally, the FWHM of these phonon modes shows distinct hysteresis, with all three modes exhibiting broader linewidths during cooling. This behaviour indicates incomplete structural recovery following the transition at $T_{C1}$. The observed thermal hysteresis, with a width $\Delta T \sim$ 100 K, strongly supports the first-order nature of the structural phase transition [38,39].

**3.4 Quasielastic scattering**

Quasielastic Raman scattering serves as a sensitive probe for exploring low-energy excitations, critical fluctuations, and relaxational dynamics that are often accompany structural and electronic phase transitions [40-42]. To gain a comprehensive insight into these dynamics within the low-frequency regime, we performed detailed Raman measurements across a range of temperatures during both heating and cooling cycles. As shown in Fig. 5(a), quasielastic scattering features prominently in the spectral region below 100 cm⁻¹, underscoring its relevance to the evolving phase behaviour of the system.

The temperature-dependent vibrational dynamics of CCPS reveal a complex interplay between displacive and order-disorder mechanisms during its structural phase transition at $T_{C1}$. Our spectroscopic analysis shows a pronounced redshift in low-frequency soft phonon modes with increasing temperature. As the system nears $T_{C1}$, these modes undergo significant damping, eventually merging into a broadened quasielastic feature. This phenomenon aligns with the overdamping of the soft mode, where its spectral weight shifts from a distinct inelastic peak to a central component centred at zero energy transfer which is called central mode (name is coming from inelastic light scattering experiments often used in ferroelectric phase transitions) [20,43]. The coupling between the soft mode and the central mode intensifies sharply near the transition temperature, reflecting the emergence of an intrinsic order-disorder component within the soft-mode fluctuations [43]. Such dynamics closely resemble those observed in canonical systems like $SrTiO_3$ [44], $Gd_2(MoO_4)_3$ [20], $Sn_2P_2Se_6$ [43], SbSI [45], $Pb_5Ge_3O_{11}$ [46], $AgNbO_3$ [47]. This behaviour of the central peak is frequently associated with the internal relaxation processes of the crystal, which are triggered by interactions with dynamic lattice defects [48], phonon density, entropy fluctuations [49,50], and the motion of domain walls [51,52].

Above the transition temperature T > $T_{C1}$, the soft phonon mode progressively fades into the tail of the quasielastic scattering background. Its intensity and linewidth exhibit notable temperature dependence across both heating and cooling cycles. The emergence of a pronounced quasielastic peak (central peak) centred at ω = 0 in the Raman response, $\chi''(\omega)$, signals the presence of critical fluctuations associated with the phase transition. Figure 5(a) presents the Raman response, defined as the imaginary part of the dynamic susceptibility, which captures the collective dynamics of the underlying excitations of the system. This response, $\chi''(\omega) \propto \frac{I(\omega)}{[n(\omega)+1]}$, is computed from the raw Raman intensity, $I(\omega)$, corrected by the Bose thermal factor, $n(\omega)$, and plotted as a function of frequency for various temperatures. Figure 5(a) shows the Raman response at different temperatures for the heating and cooling cycle in the spectral range of ~ 14-100 cm$^{-1}$. It reveals substantial changes in Raman intensity with temperature, particularly below ~ 100 cm$^{-1}$ (12.4 meV), where the low-energy features become increasingly prominent upon cooling. Correspondingly, the Raman conductivity exhibits a sharp peak near ω = 0, as shown in Figure 5(b), further confirming the development of central-mode dynamics near the structural phase boundary.

To quantitatively assess the low-energy dynamics, we focus on the temperature-dependent dynamic Raman susceptibility, $\chi^{dyn}$ (T), which is extracted from the finite-frequency Raman response using the Kramers-Kronig relation as $\chi^{dyn} = \lim_{\omega \to 0} \chi(k=0,\omega) \equiv \frac{2}{\pi} \int_0^{\Omega} \frac{\chi''(\omega)}{\omega} d\omega$ [36]. Specifically, $\chi^{dyn}$ (T) was obtained by integrating the Raman response up to Ω = 100 cm$^{-1}$ (12.4 meV), after subtracting the phonon contribution. As shown in Fig. 5(c), $\chi^{dyn}$ (T) exhibits pronounced hysteresis between heating and cooling cycles, alongside distinct transitions. During warming from 5 K, $\chi^{dyn}$ (T) decreases relatively by ~ 31% [($\chi^{dyn}$(35 K)- $\chi^{dyn}$(5 K))/$\chi^{dyn}$(5 K)*100] between 5 K and 35 K. Beyond 35 K, coinciding with the $T_N$ ~ 32 K, it increases sharply, reaching a peak at ~ 120 K with a net rise of ~ 200% [($\chi^{dyn}$(120 K)- $\chi^{dyn}$(35 K))/$\chi^{dyn}$(35 K)*100]. Above 120 K, $\chi^{dyn}$ (T) again decreases relatively by ~ 56% [($\chi^{dyn}$(300 K)- $\chi^{dyn}$(120 K))/$\chi^{dyn}$(120 K)*100] as the temperature reaches 300 K. A similar trend is observed during cooling from 300 K to 120 K, $\chi^{dyn}$ (T) increases by ~ 164%, followed by a ~ 68% decrease between 120 K and 20 K. A marginal increase is observed from 20 K to 5 K. Notably, the heating cycle reveals an AFM transition near ~ 35 K, whereas the cooling cycle shows this transition closer to 20 K, indicating a thermal delay of ~ 15 K. This hysteretic behaviour underscores the first-order nature of the transition and reflects the complex interplay between spin, lattice, and relaxational dynamics.

The temperature derivative of the dynamic Raman susceptibility, $\chi^{dyn}$ (T), provides valuable insight into the onset, progression, and completion of the structural transformation, particularly highlighting the temperature at which the transformation rate is maximal. Figure 5(d) presents the derivative curves of $\chi^{dyn}$ (T) for both heating and cooling cycles. The derivative curve showed a peak at ~ 90 K during heating, which corresponds to the maximum rate of transformation from the antipolar phase to the quasi-antipolar phase, whereas for the cooling process, the peak shifted to 70 K. At approximately 140 K, a significant dip is observed in the derivative curve. Beyond ~ 190 K, the derivative exhibits a linear behaviour, indicating a fully transformed phase, i.e., a paraelectric phase [53].

## 4. Conclusion

We present an in-depth temperature and polarization resolved Raman spectroscopic study of the multiferroic compound CCPS. Our study systematically examines the vibrational dynamics of both low- and higher-symmetry phonon modes during heating and cooling cycles. A pronounced thermal hysteresis is observed in the phonon modes below $T_{C1}$ strongly supports the first-order character of the structural phase transition. Furthermore, anomalies in the self-energy parameters of the phonon modes serve as clear signatures of multiple underlying transitions, including antiferromagnetic, antipolar, and quasi-antipolar phases. Our measurements reveal significant damping of the soft phonon mode, which ultimately evolves into a central peak associated with quasielastic scattering which is indicative of critical fluctuations. The temperature-dependent dynamic Raman susceptibility also exhibits marked hysteresis and distinct features corresponding to the phase transitions. This study resolves longstanding ambiguities surrounding the origin and evolution of the central peak and provides evidence for a crossover in the transition mechanism from a displacive to an order-disorder type.

**Data availability statement**

The data underlying this study are available in the published article and its Supporting Information.

**Acknowledgements**

P.K. thanks SERB (Project no. CRG/2023/002069) for the financial support and IIT Mandi for the experimental facilities.

**Figures:**

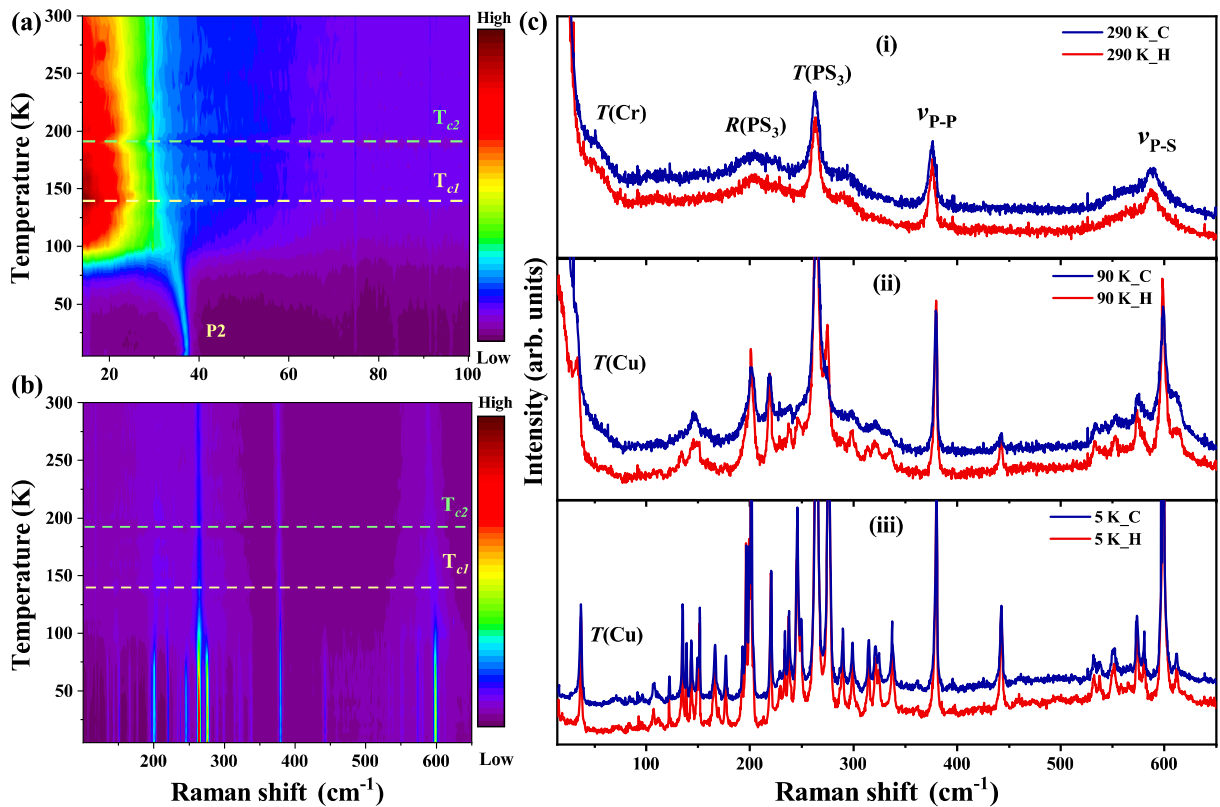

**Figure 1:** (a) Low frequency 2D colour contour plot of CCPS in the temperature range 5-300 K. (b) High frequency 2D colour contour plot in the temperature range 5-300 K. Dotted green and yellow line represents the first and second order phase transition temperature respectively. (c)(i-iii) Temperature dependent Raman spectra of CCPS measured upon heating (H -red colour) and cooling (C - blue colour) cycle at temperature 290 K, 90 K, and 5 K, respectively.

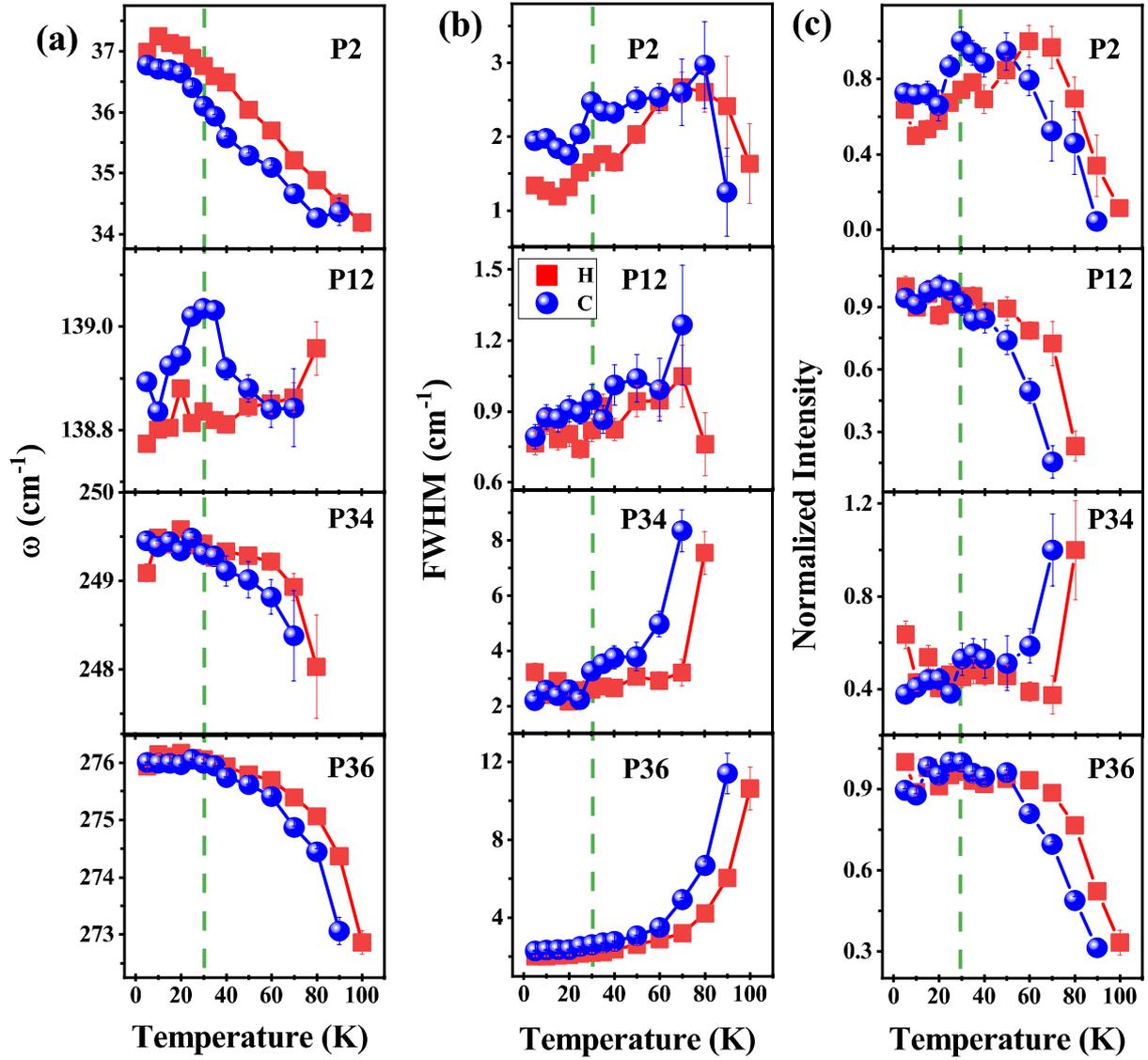

**Figure 2:** Thermal hysteresis behaviour of (a) frequency (ω), (b) FWHM, and (c) normalized intensity of the phonon modes P2, P12, P34, and P36. Blue sphere and red squares indicate cooling and heating cycle of the measurements. The dashed green line indicates the antiferromagnetic transition at $T_N \sim 30$ K.

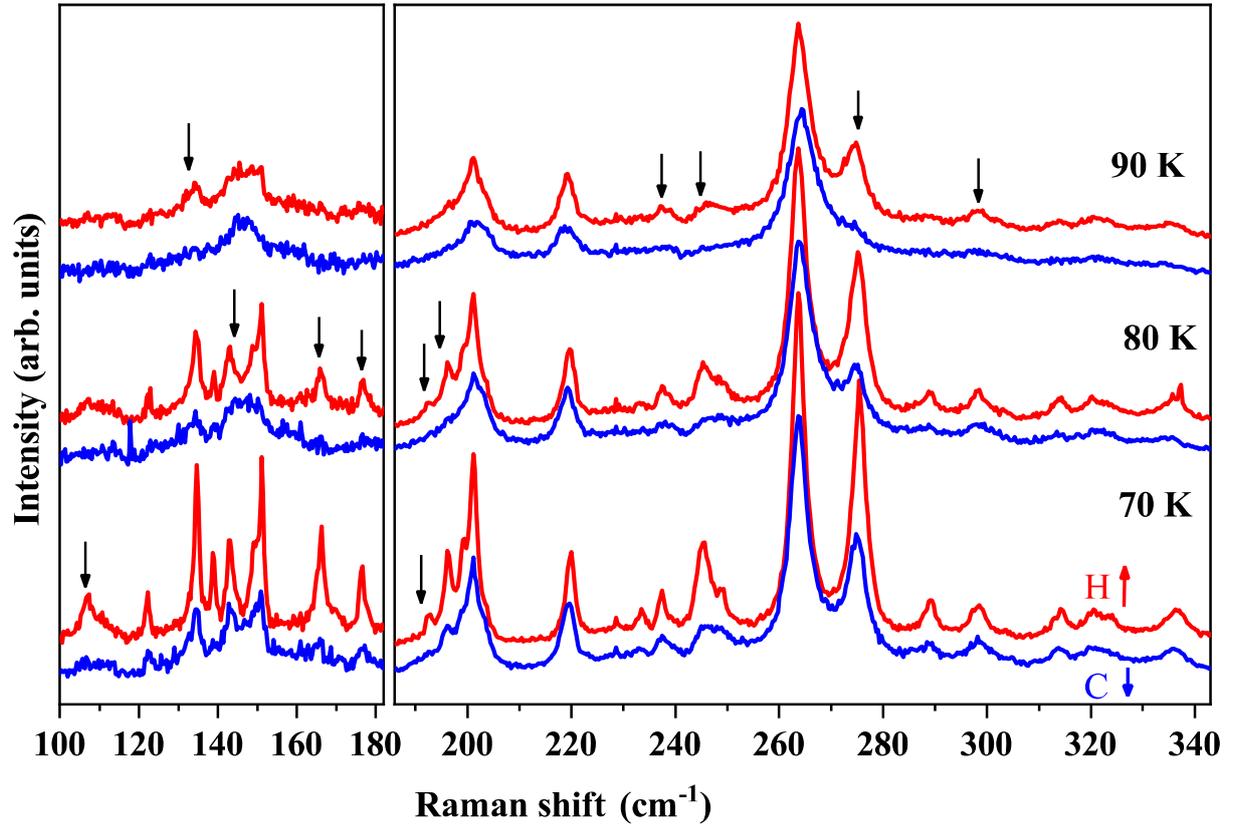

**Figure 3:** Temperature dependent Raman spectra of heating (red colour) and cooling (blue colour) cycle at 70 K, 80 K, and 90 K. Spectra recorded in the spectral range of 100-182 cm$^{-1}$ and 185-345 cm$^{-1}$ is shown. Black coloured down arrows represent the unrecovered phonon modes in cooling cycle.

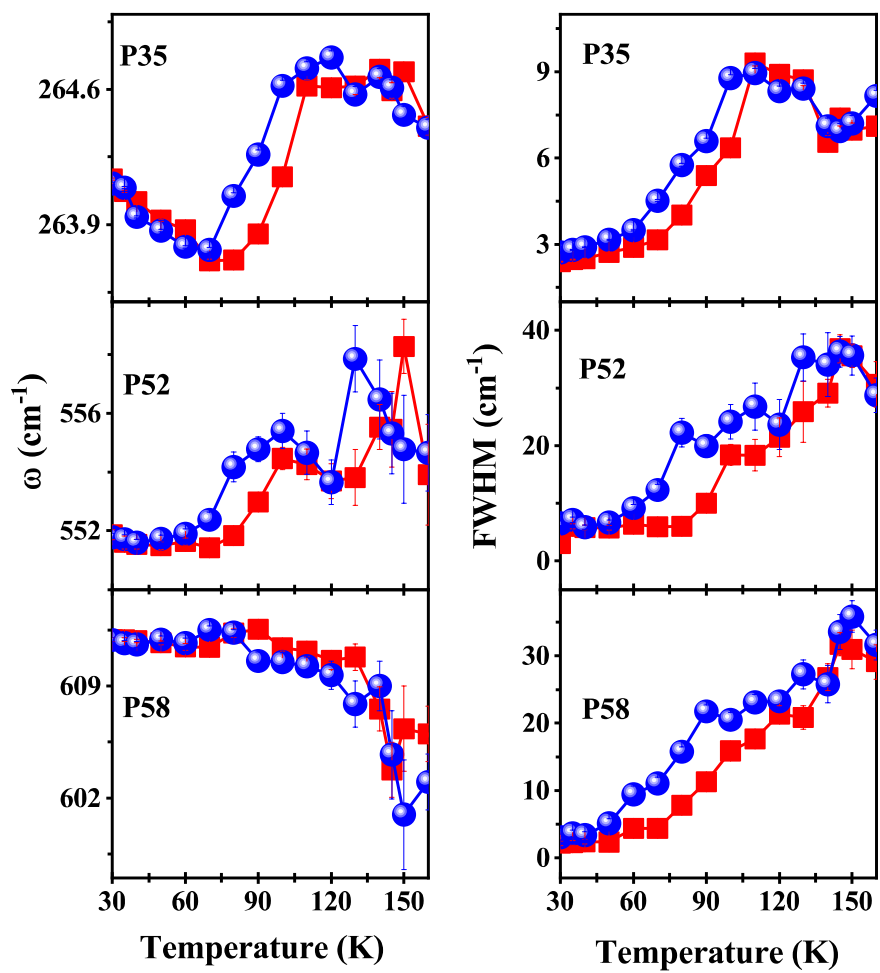

**Figure 4:** Thermal hysteresis behaviour of frequency (ω) and FWHM of the phonon modes P35, P52, and P58. Blue sphere and red squares indicate cooling and heating cycle of the measurements.

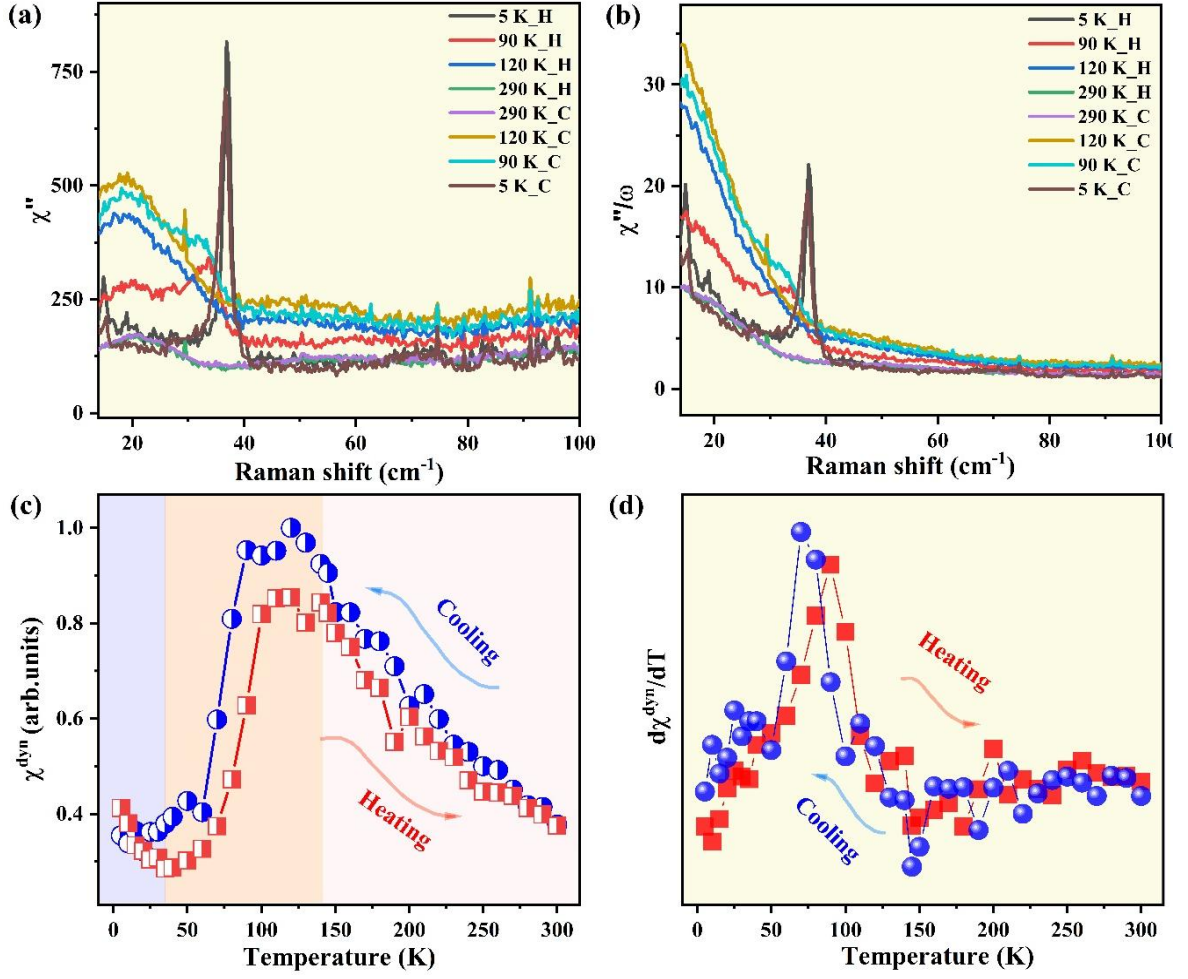

**Figure 5:** (a) Temperature evolution of the Raman response $\chi''(\omega)$, (b) Temperature dependence of the Raman conductivity $\chi''/\omega$ in the temperature range of 5 to 290 K in heating (red colour) and cooling (blue colour) cycle. (c) Temperature dependence of the dynamic Raman susceptibility $\chi^{dyn}$ (T). $\chi^{dyn}$ (T) for heating and cooling cycle is normalised w.r.t the maximum value of cooling cycle. (d) First derivative of dynamic Raman susceptibility $d\chi^{dyn}/dT$ in the temperature range of 5 to 300 K in heating and cooling cycle.

# Supplementary Material

# Hysteretic Phonons and Quasielastic Response: A Raman Study of Thermal Memory in Two-dimensional CuCrP$_2$S$_6$

Chaitanya B. Auti[1,*], Atul G. Chakkar[1], Sebastian Selter[2], Yuliia Shemerliuk[2], Bernd Büchner[2,3], Saicharan Aswartham[2], and Pradeep Kumar[1#]

[1]School of Physical Sciences, Indian Institute of Technology, Mandi-175005, India.
[2]Institute for Solid State Research, Leibniz IFW Dresden, Helmholtzstr. 20, 01069 Dresden, Germany
[3]Institute of Solid State and Materials Physics and Würzburg-Dresden Cluster of Excellence ct.qmat, Technische Universität Dresden, 01062 Dresden, Germany

## 1. Polarization dependence of the phonon modes

Polarization-dependent Raman spectroscopy is a powerful tool to probe symmetry of the phonon modes. By analyzing the angular dependence of Raman intensities under different incident and scattered light polarizations, one can determine the irreducible representations of the vibrational modes. To analyse symmetry of the phonon modes, we conducted detailed polarisation-dependent Raman measurements in parallel polarisation, i.e., incident and scattered light are parallel to each other, and cross-polarisation configuration, i.e., incident and scattered light are perpendicular to each other at 5 K and 300 K, as illustrated in Fig. S6 (a-b) and Fig. S7 (a-b). The variation in intensity as a function of polar angle can be analysed through a semi-classical approach. The Raman scattering intensity ($I$) is represented as $I \alpha |\hat{e}_s^T . R . \hat{e}_i|^2$, where $T$ denotes the transpose of the vector and R represents the Raman tensor (1). $\hat{e}_i$ and $\hat{e}_s$ denote unit vectors indicating the directions of incoming and dispersed light, respectively. In the specified configuration, the unit vectors for incident and scattered light polarisation are denoted as $\hat{e}_i = [\cos\theta \quad \sin\theta \quad 0]$ and $\hat{e}_s = [\cos\theta \quad \sin\theta \quad 0]$ for parallel configurations and $\hat{e}_i = [\cos\theta \quad \sin\theta \quad 0]$, $\hat{e}_s = [-\sin\theta \quad \cos\theta \quad 0]$ for perpendicular configurations. Here, $\theta$ represents the relative angle between $\hat{e}_i$ and $\hat{e}_s$. The Raman tensors for the phonon modes for the low-temperature $Pc$ space group ($A'$ and $A''$ symmetries) and the high-temperature $C2/c$ space group ($A_g$ and $B_g$ symmetries) given as: $R(A'/A_g) = \begin{pmatrix} b & 0 & d \\ 0 & c & 0 \\ d & 0 & a \end{pmatrix}$ and

$R(A''/B_g) = \begin{pmatrix} 0 & f & 0 \\ f & 0 & e \\ 0 & e & 0 \end{pmatrix}$. The angular dependence of the intensity of these modes for parallel and perpendicular configuration is given as:

$$I^{\parallel}_{A'(A_g)} = b^2 \cos^4\theta + c^2 \sin^4\theta + 2bc \cos^2\theta \sin^2\theta, \quad (1)$$

$$I^{\parallel}_{A''(B_g)} = f^2 \sin^2(2\theta), \quad (2)$$

$$I^{\perp}_{A'(A_g)} = \cos^2\theta \sin^2\theta (c-b)^2, \quad (3)$$

$$I^{\perp}_{A''(B_g)} = f^2 \cos^2(2\theta), \quad (4)$$

From Fig. S6 (a-b) and S7 (a-b) we observed two fold symmetry for parallel configuration and four-fold symmetries for perpendicular configuration for all observed phonon modes at 5 K and 300 K.

**Figures:**

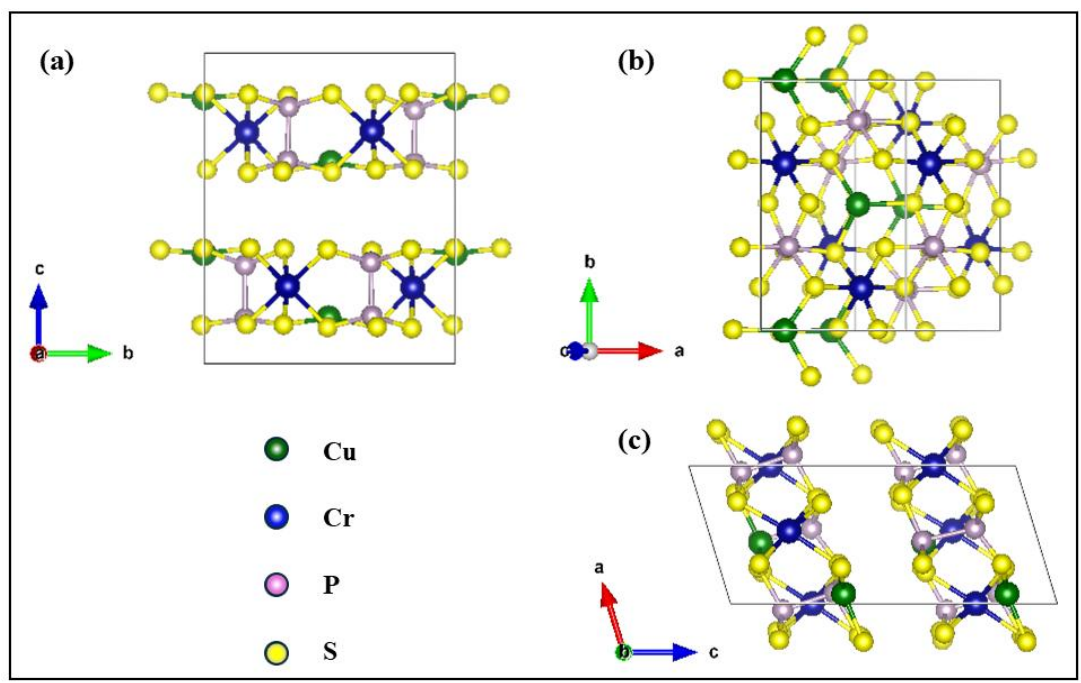

**Figure S1**: Crystal structure at low temperature *Pc* phase in the (a) *bc*, (b) *ab*, and (c) *ac* plane, respectively.

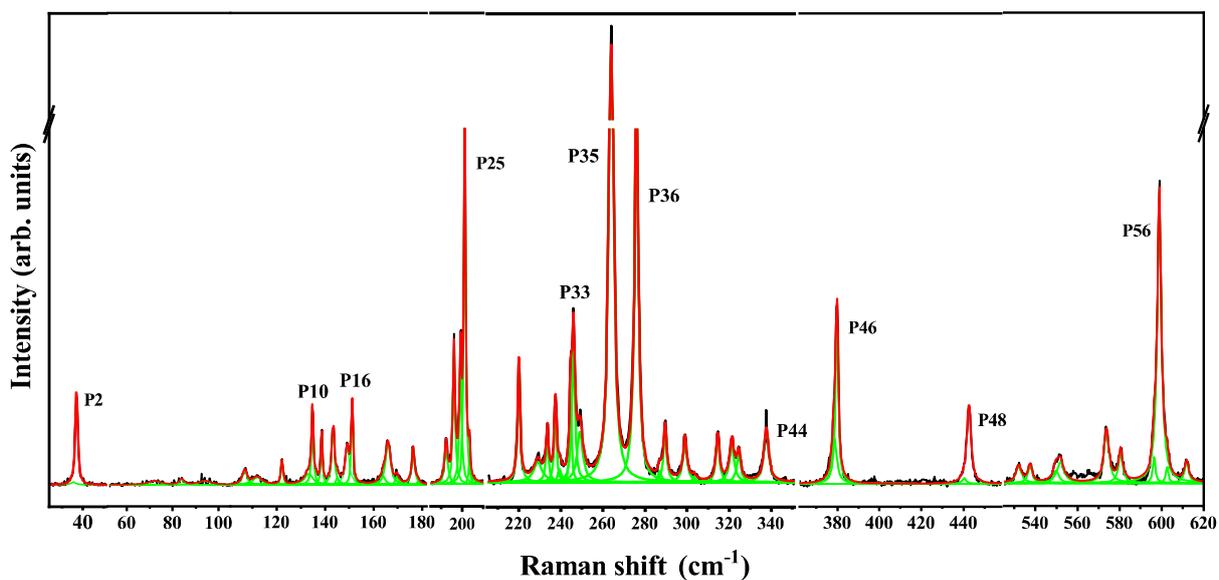

**Figure S2:** The fitted Raman spectrum of the single crystal of CCPS in the spectral range of 14-700 cm$^{-1}$ at 5 K. The phonon modes observed are designated as P1-P58. The solid thin green line represents the individual fit of the phonon modes, while a solid red line denotes the total fit of the Raman spectra.

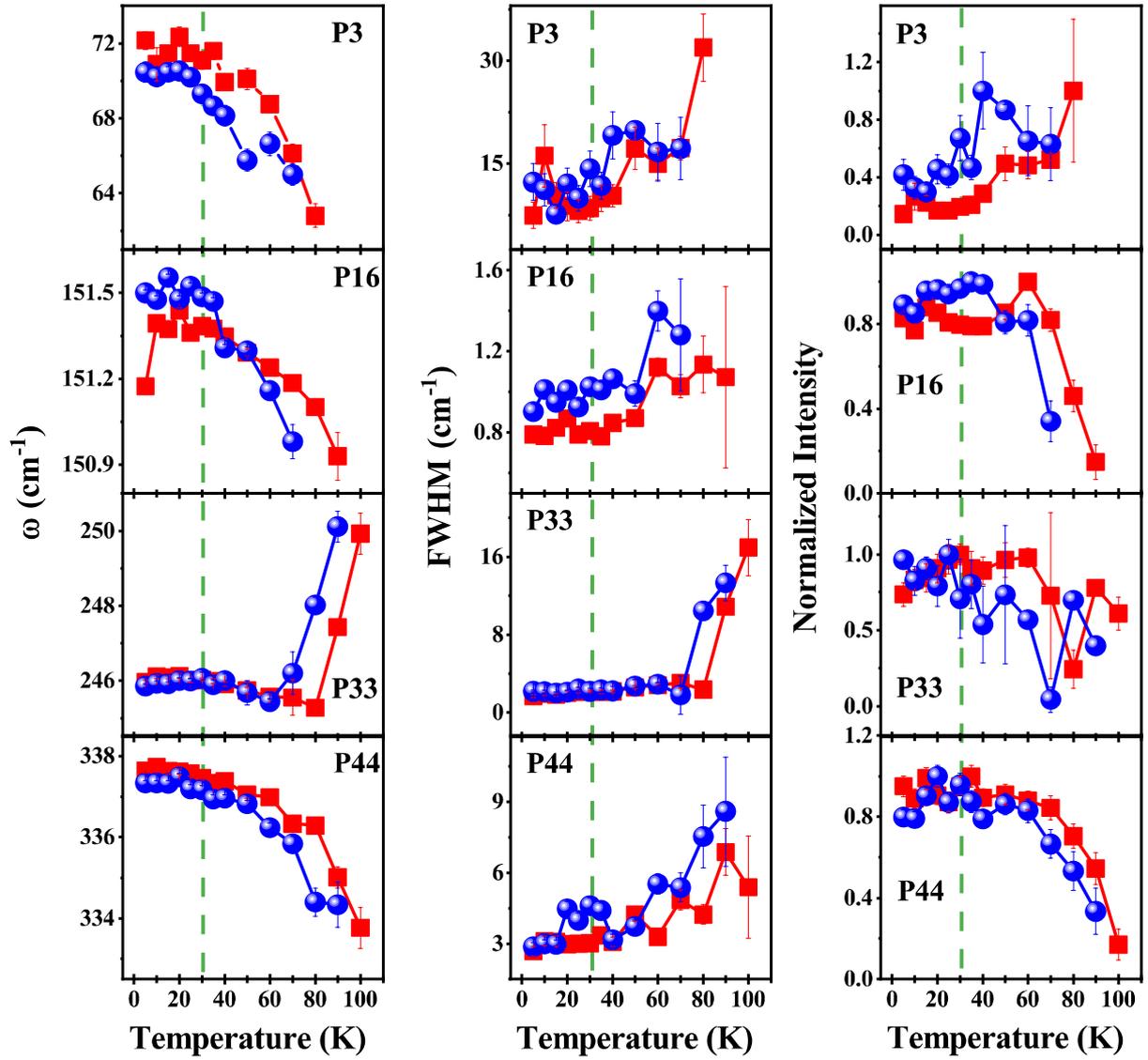

**Figure S3:** Thermal hysteresis behaviour of (a) frequency (ω), (b) FWHM, and (c) normalized intensity of the phonon modes P3, P16, P33, and P44. Blue spheres and red squares indicate cooling and heating cycle of the measurements. The dashed green line indicates the antiferromagnetic transition at $T_N \sim 30$ K.

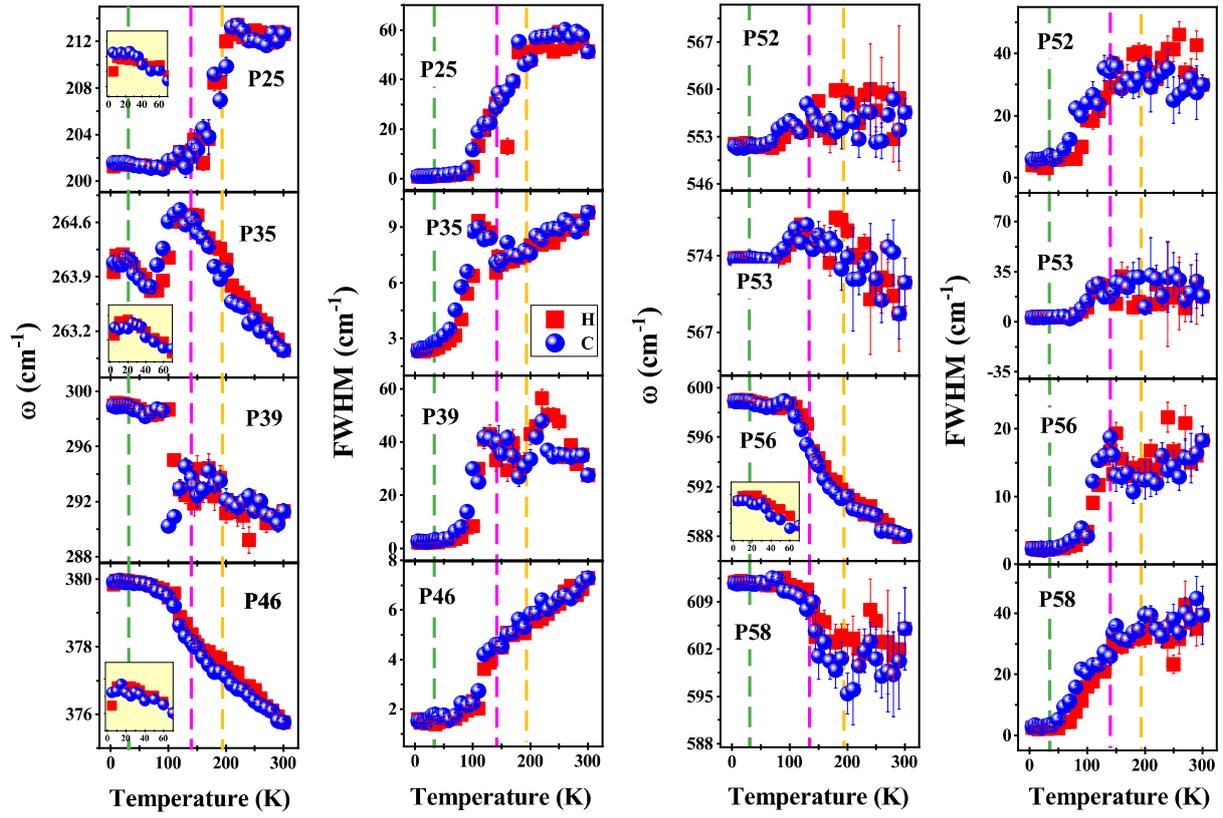

**Figure S4:** Thermal hysteresis behaviour of mode frequency (ω), and linewidth (FWHM) of phonon modes P25, P35, P39, P46, P52, P53, P56, and P58. The dashed green line indicates the antiferromagnetic transition at $T_N$ ~ 30 K. The dashed pink, and orange lines shows at antipolar and quasiantipolar phase transition ~ 140 K and ~ 190 K, respectively.

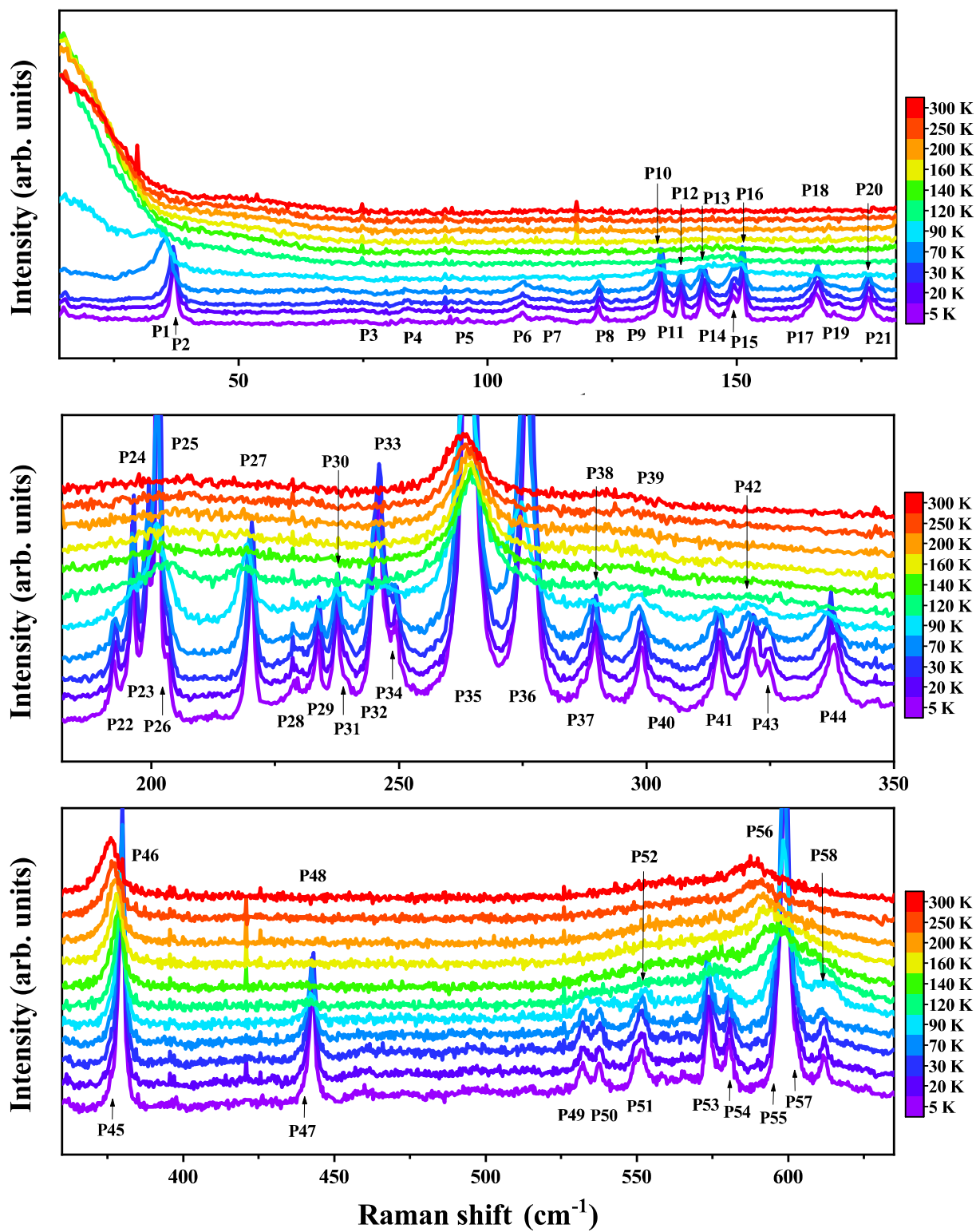

**Figure S5.** Temperature evolution of the Raman spectrum of CCPS in the frequency range of 14-180 cm$^{-1}$, 180-350 cm$^{-1}$, 350-680 cm$^{-1}$.

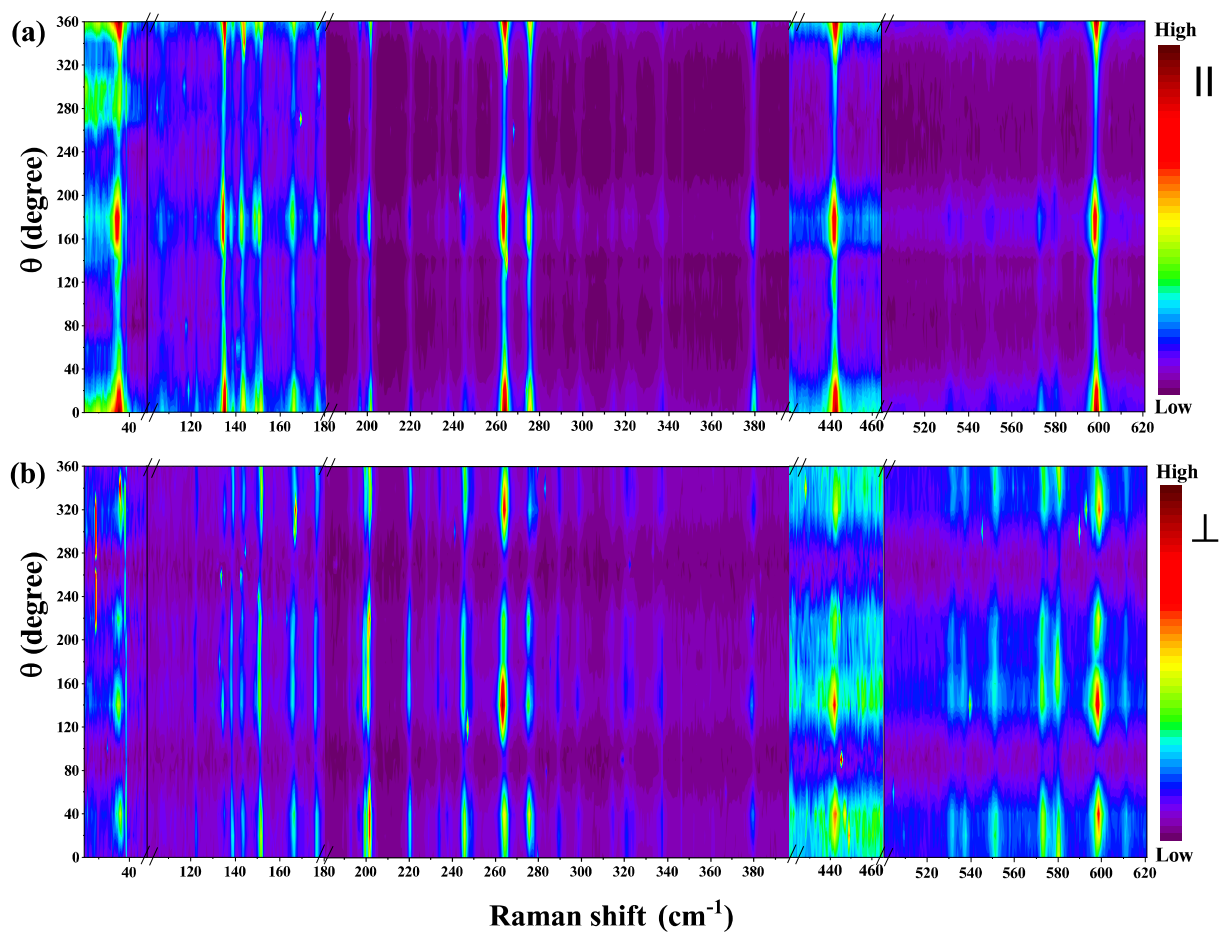

**Figure S6:** 2D colour contour map of polarization-dependent intensity of the phonon modes measured at 5 K in (a) Parallel configuration and (b) Perpendicular configuration.

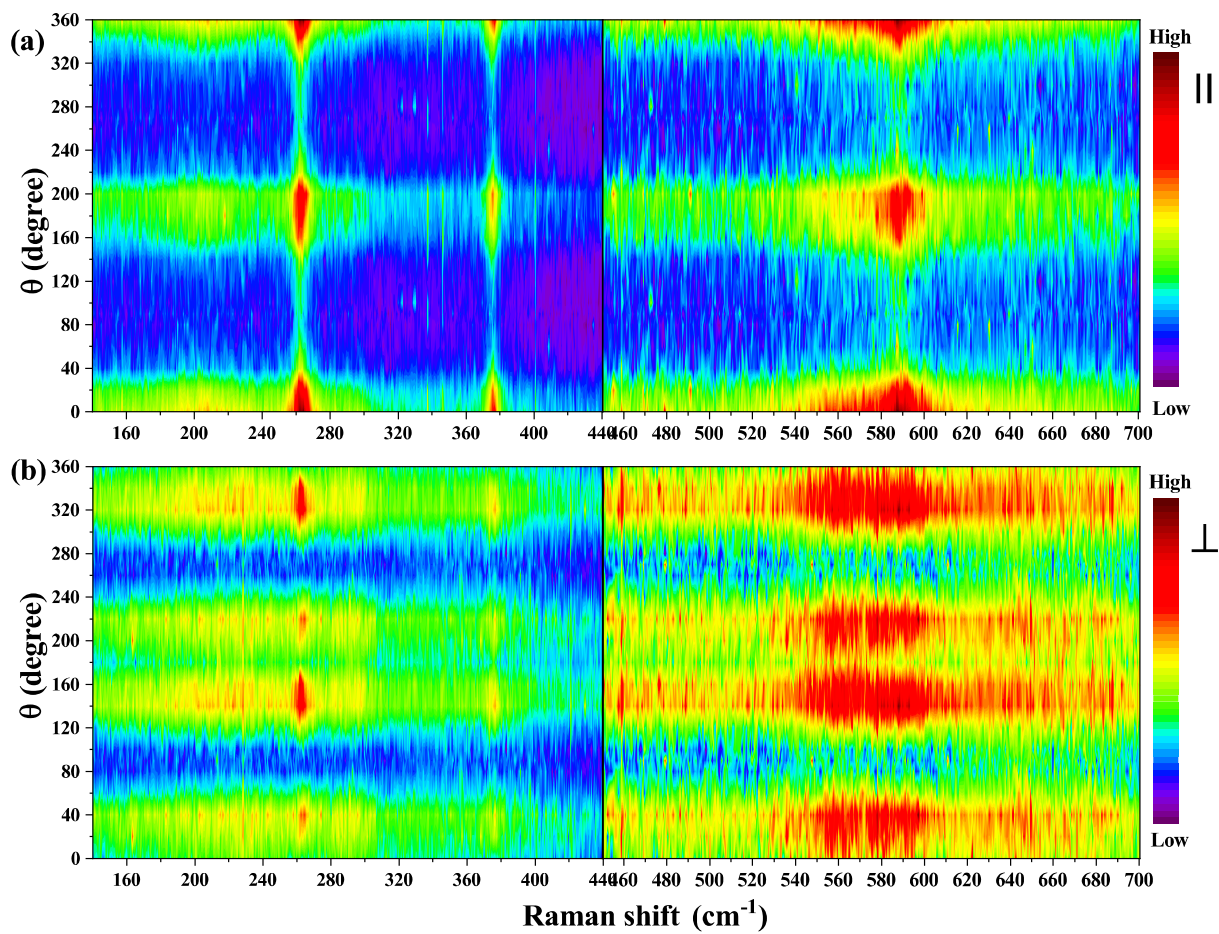

**Figure S7:** 2D colour contour map of polarization-dependent intensity of the phonon modes measured at 300 K in (a) Parallel configuration and (b) Perpendicular configuration.

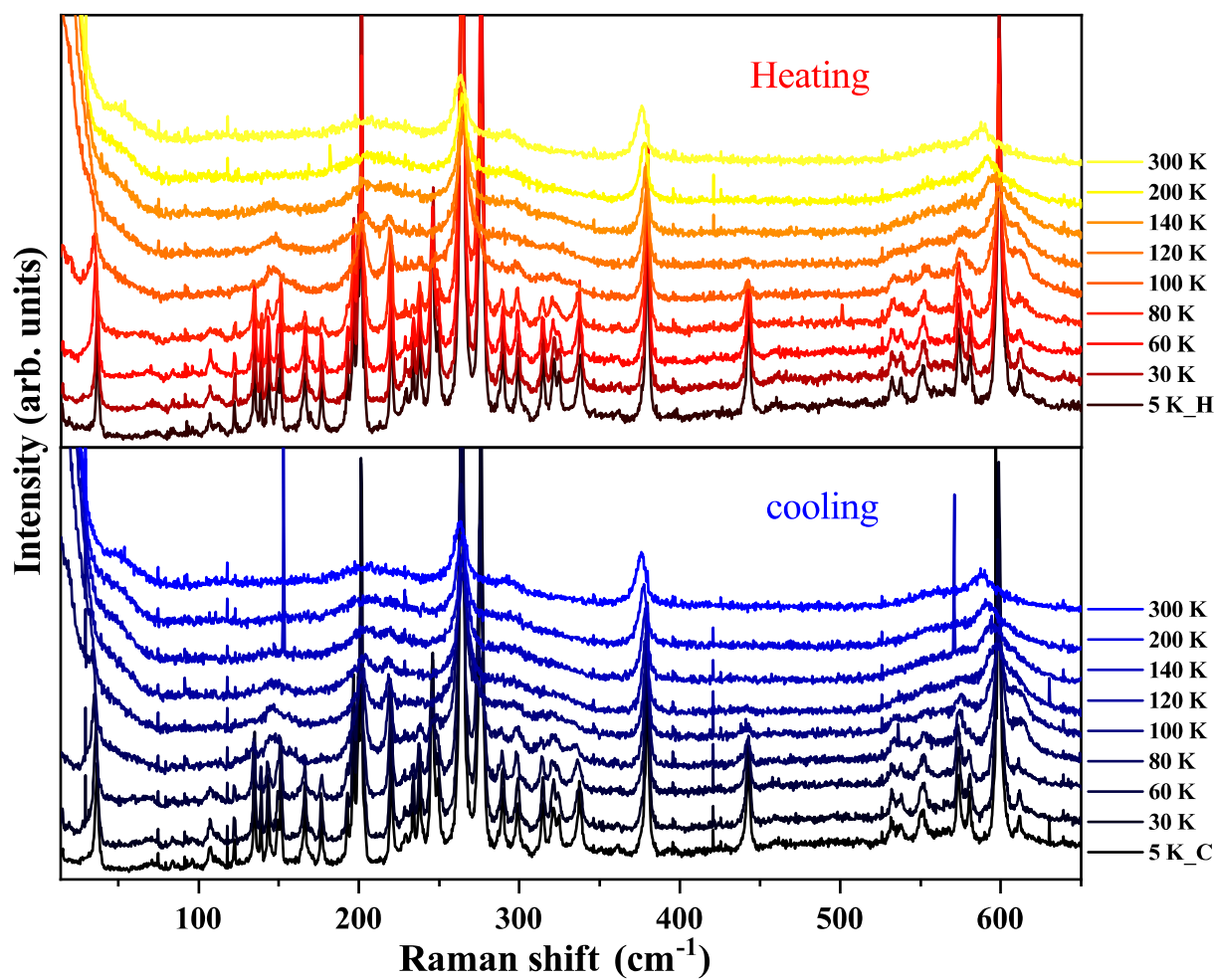

**Figure S8:** Temperature dependent Raman spectra of CCPS measured upon heating (H) and cooling (C) cycle represented by red and blue spectrum respectively.

**Tables:**

**Table 1.** List of the experimentally observed modes with their frequencies at 5 K and 300 K in first heating cycle. Units are in cm$^{-1}$.

| Modes | ω (5 K) | ω (300 K) | Modes | ω (5 K) | ω (300 K) |
|---|---|---|---|---|---|
| P1 | 35.5 ± 1.07 | - | P30 | 239.4 ± 0.29 | - |
| P2 | 36.9 ± 0.01 | - | P31 | 237.4 ± 0.03 | - |
| P3 | 72.1 ± 0.49 | - | P32 | 244.7 ± 0.04 | - |
| P4 | 83.8 ± 0.20 | - | P33 | 245.9 ± 0.03 | - |
| P5 | 94.0 ± 0.47 | - | P34 | 249.0 ± 0.10 | - |
| P6 | 106.8 ± 0.09 | - | P35 | 263.9 ± 0.00 | 262.9 ± 0.05 |
| P7 | 111.9 ± 0.28 | - | P36 | 275.9 ± 0.00 | - |
| P8 | 122.1 ± 0.03 | - | P37 | 287 ± 0.26 | - |
| P9 | 133.2 ± 0.50 | - | P38 | 289.5 ± 0.05 | - |
| P10 | 134.7 ± 0.01 | - | P39 | 298.9 ± 0.06 | 291.2 ± 0.54 |
| P11 | 136.4 ± 0.23 | - | P40 | 303.4 ± 0.51 | - |
| P12 | 138.7 ± 0.01 | - | P41 | 314.5 ± 0.05 | - |
| P13 | 143.3 ± 0.01 | - | P42 | 321.2 ± 0.09 | - |
| P14 | 145.0 ± 0.14 | - | P43 | 324.5 ± 0.13 | - |
| P15 | 149.1 ± 0.03 | - | P44 | 337.6 ± 0.06 | - |
| P16 | 151.1 ± 0.01 | - | P45 | 378.5 ± 0.17 | - |
| P17 | 163.9 ± 0.29 | - | P46 | 379.8 ± 0.02 | 375.7 ± 0.06 |
| P18 | 166.0 ± 0.05 | - | P47 | 440.4 ± 0.47 | - |
| P19 | 169.7 ± 0.12 | - | P48 | 442.6 ± 0.03 | - |
| P20 | 176.4 ± 0.02 | - | P49 | 532.0 ± 0.15 | - |
| P21 | 178.4 ± 0.27 | - | P50 | 537.4 ± 0.12 | - |
| P22 | 192.4 ± 0.05 | - | P51 | 549.7 ± 0.58 | - |
| P23 | 196.1 ± 0.01 | - | P52 | 551.9 ± 0.55 | 556.6 ± 1.99 |
| P24 | 199.2 ± 0.01 | - | P53 | 573.7 ± 0.04 | 571.5 ± 2.18 |
| P25 | 201.2 ± 0.00 | 212.6 ± 0.47 | P54 | 580.4 ± 0.07 | - |
| P26 | 203.4 ± 0.04 | - | P55 | 596.4 ± 0.07 | - |
| P27 | 220.0 ± 0.01 | - | P56 | 598.8 ± 0.00 | 588.0 ± 0.33 |
| P28 | 229.1 ± 0.25 | - | P57 | 602.6 ± 0.11 | - |
| P29 | 233.5 ± 0.05 | - | P58 | 611.8 ± 0.11 | 605 ± 6.05 |